\documentclass[conference]{IEEEtran}
\usepackage{filecontents,lipsum}
\usepackage[noadjust]{cite}
\usepackage{color}
\usepackage{graphicx}   
\usepackage{array}
\usepackage{dcolumn}    
\usepackage{bm}         
\usepackage[normalem]{ulem}

\usepackage{cite}

\ifCLASSINFOpdf
\else
\fi

\usepackage{soul}
\usepackage{acronym}
\usepackage{cite}

\acrodef{EM}{electromagnetic}
\acrodef{NoC}{Network-on-Chip}

\begin{document}

\title{Intercell Wireless Communication in Software-defined Metasurfaces}

\author{
\IEEEauthorblockN{
Anna C. Tasolamprou\IEEEauthorrefmark{1}, Mohammad Sajjad Mirmoosa\IEEEauthorrefmark{2}, Odysseas Tsilipakos\IEEEauthorrefmark{1}, Alexandros Pitilakis\IEEEauthorrefmark{1}\IEEEauthorrefmark{3}, Fu Liu\IEEEauthorrefmark{2},\\
Sergi Abadal\IEEEauthorrefmark{4}, Albert Cabellos-Aparicio\IEEEauthorrefmark{4}, Eduard Alarc\'{o}n\IEEEauthorrefmark{4}, Christos Liaskos\IEEEauthorrefmark{1}, Nikolaos V. Kantartzis\IEEEauthorrefmark{1}\IEEEauthorrefmark{3},\\
Sergei Tretyakov\IEEEauthorrefmark{2}, Maria Kafesaki\IEEEauthorrefmark{1}\IEEEauthorrefmark{5}, Eleftherios N. Economou\IEEEauthorrefmark{1}, Costas M. Soukoulis\IEEEauthorrefmark{1}\IEEEauthorrefmark{6}
}
\IEEEauthorblockA{\IEEEauthorrefmark{1}Foundation for Research and Technology Hellas, 71110, Heraklion, Crete, Greece}
\IEEEauthorblockA{\IEEEauthorrefmark{2}Department of Electronics and Nanoengineering, Aalto University, P.O. Box 15500, Espoo, Finland}
\IEEEauthorblockA{\IEEEauthorrefmark{3}Department of Electrical and Computer Engineering, Aristotle University of Thessaloniki, Thessaloniki, Greece}
\IEEEauthorblockA{\IEEEauthorrefmark{4}NaNoNetworking Center in Catalonia (N3Cat), Universitat Polit\`{e}cnica de Catalunya, Barcelona, Spain}
\IEEEauthorblockA{\IEEEauthorrefmark{5} Department of Materials Science and Technology, University of Crete, 71003, Heraklion, Crete, Greece}
\IEEEauthorblockA{\IEEEauthorrefmark{6} Ames Laboratory and Department of Physics and Astronomy, Iowa State University, Ames, Iowa 50011, USA\\
Email: atasolam@iesl.forth.gr
}
}

\maketitle

\begin{abstract}

Tunable metasurfaces are ultra-thin, artificial electromagnetic components that provide engineered and externally adjustable functionalities. The programmable metasurface, the HyperSurFace, concept consists in integrating controllers within the metasurface that interact locally and communicate globally to obtain a given electromagnetic behaviour. Here, we address the design constraints introduced by both functions accommodated by the programmable metasurface, i.e., the desired metasurface operation and the unit cells wireless communication   enabling such programmable functionality. The design process for meeting both sets of specifications is thoroughly discussed. Two scenarios for wireless intercell communication are proposed. The first exploits the metasurface layer itself, while the second employs a dedicated communication layer beneath the metasurface backplane. Complexity and performance trade-offs are highlighted.

\end{abstract}


%
\IEEEpeerreviewmaketitle

\section{Introduction}
Metasurfaces  are planar artificial structures which have recently enabled the realization of novel, ultra-thin \ac{EM} components with engineered response \cite{Soukoulis:2011,Glybovski2016}. An abundance of functionalities has been demonstrated \cite{chen2016review,Hsiao:2017}, including perfect absorption or wavefront manipulation. Obviously, tunability or reconfigurability are highly desirable in this context. Initial studies revolved around achieving global tunability by means of external stimuli (heat, voltage, light) \cite{chen2016review}. To add reconfigurability and the ability to host multiple functionalities, recent works have integrated biased diodes within each unit cell so that the response of each unit cell can be tuned locally \cite{Cui2014,Yang:2016}.


A step further towards the compelling vision of interconnectable, fully adaptive metasurfaces with multiple concurrent functionalities is the concept of HyperSurFace (HSF) \cite{Liaskos2015}. The HSF paradigm builds upon the description of \ac{EM} functionalities in reusable software modules. Such \emph{software-defined} approach allows authorized users to easily change the behavior of the metasurface by sending preset commands. To disseminate, interpret, and apply those commands, a HSF requires the integration of a network of miniaturized controllers within the metamaterial structure. This poses several implementation and co-integration challenges \cite{AbadalACCESS}, among which we highlight and focus on the interconnection of the internal controllers.

Communication among the controllers of a HSF can be either wired or wireless. \emph{A priori,} wired means are preferable as the interconnect will be most likely co-integrated with the controllers within the same chip \cite{AbadalACCESS} and because knowledge from similar scenarios like low-power embedded systems can be reused \cite{Bjerregaard2006, Bertozzi2014}. However, issues may appear when scaling the HSF in size or in controller density: in the former case, HSFs will contain multiple chips leading to complex layout issues related to combining on-chip and off-chip interconnects; in the latter case, HSFs will integrate very dense networks leading to higher latency and power consumption if conventional NoC topologies are used \cite{Balfour2006}.

\begin{figure}[!t]
\vspace{-0.2cm}
\centering
\includegraphics[width=80mm]{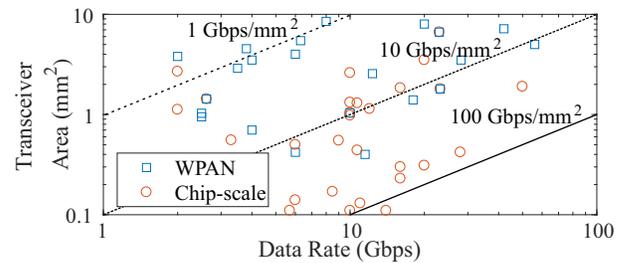}
\vspace{-0.2cm}
\caption{\label{figIntro} Area as a function of the data rate for state-of-the-art transceivers for Wireless Personal Area Networks (WPAN) and chip-scale applications. Data extracted from \cite{AbadalTON, Geng2015, Pang2017,  Tokgoz2016, Wu2016, Okada2014, Dolatsha2017, Siligaris2011, Byeon2013, Lee2015, Fujishima2013, Foulon2014, Fritsche2017, Moghadami2015, Sarmah2016, Kang2015, Thyagarajan2015, Nakajima2015, Wang2014trans, Yu2014, Weissman2016, Yang2014, Kim2017} and references therein.}
\vspace{-0.2cm}
\end{figure}


Wireless intercell communication becomes a compelling alternative in either large or dense HSFs. The use of a shared medium allows to reduce the latency and power of collective and long-range communications used during command dissemination. Also, the lack of wiring between nodes facilitates off-chip and even off-{HSF} communication. This approach is possible due to recent advances in on-chip antennas in mmWave and THz bands \cite{Markish2015, Gutierrez2009, Cheema2013}, as well as the constant miniaturization of RF transceivers for short-range applications. As shown in Fig. \ref{figIntro}, transceivers with multi-Gbps speeds and footprints as small as 0.1 mm\textsuperscript{2} have been demonstrated.

Before assessing the potential applicability of existing transceivers, it is crucial to understand the \ac{EM} propagation within this new enclosed and monolithic scenario. Some works have studied propagation in applications with metallic enclosures, but provided little room for co-design \cite{Genender2010, Ohira2011, Khademi2015}. Others have investigated propagation within a computing package \cite{Branch2005, Zhang2007, Kim2016mother}, but the structure differed considerably from HSFs.

This paper performs, for the first time, a study towards the characterization of the wireless channel within a software-defined HSF. To this end, we describe two possible \ac{EM} propagation paths in Sect. \ref{sec:envDes}, namely, through the metasurface layer or in a dedicated waveguide. We then analyze the field distribution and coupling between mmWave antennas for both cases in Sect. \ref{sec:dipdip} and \ref{sec:dedicated}. Finally, Sect. \ref{sec:conc} concludes the paper.

\section{\label{sec:envDes}Structure, environment description and electromagnetic operations}
\begin{figure}[ht]
\centering
\includegraphics[width=80mm]{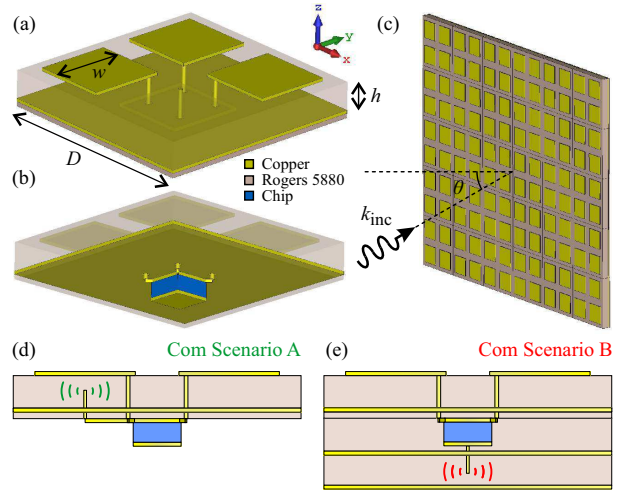}
\vspace{-0.2cm}
\caption{\label{fig1} HSF unit cell: (a) Top-view and geometric parameters, (b) bottom-view with chip for the programmable operation. (c) MS  operating at 5 GHz under oblique incidence. (d,e) Unit-cell side-view illustrating the two communication channels. (d) Scenario A:  communication in the metasurface substrate (e) Scenario B: communication  in a dedicated parallel-plate waveguide.}
\vspace{-0.2cm}
\end{figure}

\vspace{-0.1cm}
As a  case study we consider the software-defined HSF depicted in Fig.~\ref{fig1}. The metasurface (MS) part consists of an array of electromagnetically thin metallic patches placed over a dielectric substrate  back-plated by a metallic layer. To enable the software-based MS control, the patches are connected to a group of controller chips that lie below the metallic back plane through vertical vias. The controllers adjust the electromagnetic behaviour of the metasurface fabric by attributing additional local resistance and reactance at will \cite{Cui2014,Yang:2016}.  The controller plane is decoupled from the MS thanks to the back plane that separates the patches from the chips. We assume at this point that each chip serves four metallic patches. Our case study MS is designed for perfect absorption and anomalous reflection operation  in the microwave regime. For operation in the microwave regime, the size of the metasurface is required to be in the order of millimetres. Specifically the reference MS structure under consideration is designed to operate at $f=$ 5 GHz ($\lambda_0=60$~mm). It consists  of periodically arranged, four-patch unit cells with $xy$ size $D \times D = $ 12 mm  $\times$ 12 mm, as seen in Fig.~\ref{fig1}. The size of each patch is $w \times w = $ 4.2 mm $\times$ 4.2  mm. The thickness of the substrate  is  $h=1.575$ mm and it is made of Rogers RT/Duroid 5880 with permittivity  $\epsilon_r=2.2$ and loss tangent $\tan\delta=9\times10^{-4}$.	

The physical landscape of the software-defined HSF  offers several opportunities for the propagation of RF signals within the structure for wireless connectivity between the different controllers.  The actual implementation  depends on the tile lateral dimensions  and the targeted wavelength. In this work we consider two distinct communication channels, seen in Fig.~\ref{fig1}(d,e). The first channel is  the space between the MS patches and the back plane, called  MS layer (scenario A). A blind via fed form the chip serves as the antenna, while the metallic patches and the metallic back plane acts as a waveguide. The second channel is a dedicated communication plane formed by adding extra metallic plates below the chip (scenario B). Monopoles fed from the chip are inserted in the parallel-plate waveguide and  excite waves that propagate within this obstacle-free environment. Note, that in both scenarios the wave propagates in a restricted waveguide which could be considered a wire and not a free-space environment. However,  features such as the probes omnidirectional radiation, the gap leakage in scenario A and the multi-scattering in scenario B are closely related to free-space wave propagation; hence we adopt the wireless term. The selected communication path may give rise to some undesired phenomena, such as radiation losses or interference, but, on the other hand, provides enhanced design opportunities and functionalities.

To ensure that the electromagnetic response of the MS and the wireless communication operation are decoupled, we choose the communication frequency to be greater than 25 GHz. This decoupling is especially important in scenario A where the metasurface layer hosts both the electromagnetic waves  for the MS operation as well as the communication signals. Therefore, overall, we investigate the channel communication in the range $f=$ [25 GHz, 200 GHz]. The distance between two neighbouring nodes equals $D$ and is in the order of 5$\lambda$ to 40$\lambda$, respectively; this means that the communication takes place in the near and intermediate field regime. Thus, unable to resort to simplified farfield manipulation, we use full wave electromagnetic analysis for the numerical investigation. For higher frequencies, i.e., for frequencies $f>1$ THz ($D>200\lambda$) the full wave analysis becomes cumbersome and we need to turn to simplified schemes such as ray tracing \cite{Kantelis20141823}. It is stressed that even though we perform the analysis for the reference case dimensions, a direct scaling of the structure along with the wavelengths of operation is possible as long as the properties of the materials involved remain the same.

\section{\label{sec:dipdip}Cell to cell communication in the metasurface layer}
\vspace{-0.1cm}
The MS layer communication channel of the software-defined HSF is shown in Fig.~\ref{fig1}(d). For efficient communication, the electromagnetic energy should be confined between the periodic copper patches and the ground; the waves should not leak to the free-space above. This leakage is a path loss for the communication channel and should be minimized. Our study will be focused on two neighbouring unit cells with respect to the maximum power which can be transmitted from one cell to the other one. The antenna is connected to the chip and  located under the center of one of the patches through a cylindrical hole that isolates it form the back plane (ground). The height of the probe antenna is  $L=1.4$ mm. Due to the presence of the ground, this probe may operate as a quarter-wavelength monopole antenna; however, the complex environment of the MS is expected to affect the antenna operation. The corresponding frequency is $f_0={c_0/(4L\sqrt{\varepsilon_{\rm{r}}}})  \approx36$ GHz. Ideally, i.e., in the absence of the copper patches, an antenna resonance (zero reactance) is expected at this frequency. The waveguide port feeding this antenna was designed to match the theoretical $\lambda/4$ monopole input impedance, without any additional optimization for the actual structure.

We evaluate the neighbouring nodes communication by calculating the corresponding scattering matrix. To ensure that the MS  and the communication operations are electromagnetically decoupled, we assume that the frequency is greater than 30 GHz (the MS operational frequency is 5 GHz). To minimize the free-space leakage, the gap between the patches, $w_{gap}$, should be electromagnetically small. In our case study the gap is equal to $w_{gap}=1.8$ mm, therefore the absolute upper bound in the studied frequency range should not exceed $100$~GHz ($\lambda_0=$3 mm).
Hence, we simulate our structure in the frequency range $f$=[30 GHz,100 GHz]. We employ ANSYS HFSS, a commercial, 3D full-wave simulator based on the finite element method (FEM). To evaluate the communication between the antennas we calculate the transmission coefficient $S_{21}$ (dB) which corresponds to the power fraction collected by the receiver. In addition, we calculate the reflection $S_{11}$ coefficient which reveals the power fraction reflected back to the emitter. For optimum operation the magnitude of $S_{11}$ should be low, meaning negligible reflection, and the magnitude of $S_{21}$ should be high. $S_{11}$ can be improved by employing an external matching circuit so we focus here  on $S_{21}$.
\begin{figure}[ht]\centering
	\includegraphics[width=80mm]{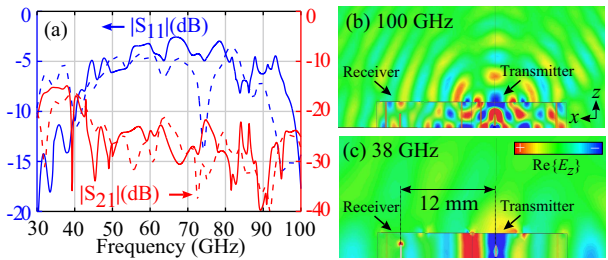}
	\vspace{-0.2cm}
	\caption{(a) Scattering components $S_{11}$ and $S_{21}$. Dashed and solid curves correspond to the initial and optimized structure, respectively. $E_z$ component at (b) $f=$~100~GHz and (c) $f=$~38~GHz for the optimized structure.}
	\label{fig:smatrix}
\vspace{-0.2cm}
\end{figure}

Figure~\ref{fig:smatrix}(a) presents the transmission and reflection coefficients for the present structure under study (dashed lines). As observed, $\vert S_{21}\vert$ is smaller than -20~dB after 45~GHz. However, around the frequency of 40~GHz, it is larger than -20~dB which is acceptable from the point of view of communication. $\vert S_{11}\vert$ has local minima at approximately 40~GHz, 75~GHz and 95~GHz. However, the transmission coefficient $\vert S_{21}\vert$  is maximum at 40~GHz. Thus, we adopt this frequency for wireless intercell communication. Since the low reflection coefficient does not necessarily correspond to a high transmission coefficient, we focus on the environment
effect and the free-space leakage.  This can be seen in Fig.~\ref{fig:smatrix} by comparing the S-parameters at 40~GHz and, at~75 GHz and~95 GHz. At the high frequencies, 75~GHz and 95~GHz, the low-reflected wave radiates into the free space rather than coupling to the receiving antenna. 
To improve the communication between the transmitting and receiving antennas, we optimize the geometry parameters of the structure. We keep in mind that any geometrical modification is going to affect the MS operation, shifting the resonance frequency at higher or lower values. However, if the modifications are moderate, we can readjust the MS resonance at 5 GHz by  tuning the resistance and reactance values of the chip. The way to minimize the free-space leakage is by decreasing the gap between the patches $w_{gap}$. Additionally we can increase the substrate thickness. Finally we select the modified parameters that optimize the communication operation; the optimum patch gap is $w_{gap}^{opt}=1$ mm and the optimum thickness is $h^{opt}=2.6$ mm. The corresponding S-parameters are shown as solid lines in Fig.~\ref{fig:smatrix}(a). As can be seen, $\vert S_{21}\vert$ is significantly improved in the range $f=$[30 GHz, 40 GHz] (the local maximum is now -15 dB). Notice that at the same frequency range the reflection coefficient is also improved compared to the initial structure. Above 40 GHz the communication efficiency decreases, similarly to the initial structure, but remains, on average, higher than before. The distribution of the electric field $E_z$ is shown in Fig.~\ref{fig:smatrix}(b) and Fig.~\ref{fig:smatrix}(c) at  frequencies  $f=100$ GHz and $f=38$ GHz, respectively. At $f=100$ GHz there is significant leakage whereas at $f=38$ GHz the field is confined within the MS layer. This agrees with the increased $\vert S_{21}\vert$ coefficient at $f=38$ GHz.

\section{\label{sec:dedicated}Communication in a dedicated parallel plate waveguide}
\vspace{-0.1cm}

In this scenario we consider that the communication in the software-defined HSF is enabled by an additional channel,  dedicated solely to transferring the signals between the communication nodes,  Fig. \ref{fig1}(e). The channel is created by introducing an additional metallic plate behind the chip backplane at a distance that, as explained, is specified by the desired frequency of operation. We assume that the space between the two metallic plates is empty (air). The two metallic plates and the uniform dielectric space between them, form a parallel-plate waveguide. Each node consists of a probe antenna connected to the chip through a vertical small hole in the ground plane, as seen in Fig. \ref{fig1}(e). The communication channel is totally electromagnetically isolated form the MS layer, thus all coupling is excluded. Moreover, the parallel plates create a closed space where no energy leakage is allowed (the holes are electromagnetically small). For these reasons, this option offers robustness and design flexibility.

The parallel-plate waveguide sustains the propagation of TEM (Transverse ElectroMagnetic) waves in which both the electric and magnetic fields are perpendicular to the propagation direction. The TEM mode can be excited from zero frequency (DC) and is the only propagation mode supported by the waveguide up to the cut-off frequency of the first higher-order mode: $f<{c_0}/({2d})$. The probe acts as an omnidirectional antenna that transmits or receives electromagnetic energy omnidirectionally in the horizontal plane $xy$. In the vertical plane, the radiation is confined by the metallic plates.
\begin{figure}[ht]\centering
\includegraphics[width=80mm]{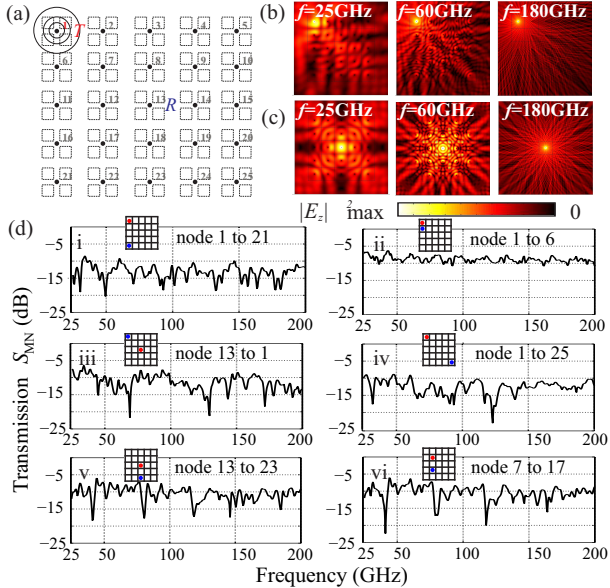}
\vspace{-0.2cm}
\caption{\label{fig3} (a) Schematic of the TEM parallel waveguide 2D approximation, node no.1 radiates and node no.13 receives. (b) and (c) Electromagnetic energy distribution at $f=$ 25GHz, $f=$ 60 GHz and $f=$ 180 GHz when the emitter is no.1 and no.13 respectively. (d) Power received at the node $M$ when node $N$ radiates, $S_{MN}$, over the frequency range $f$ =[25 GHz, 200 GHz]. Six cases of $MN$ node pairs are schematically depicted in the insets. }
\vspace{-0.2cm}
\end{figure}

Since the EM energy is carried by the single TEM mode, the waveguide is naturally impedance matched with free-space; this allows  the following approximation: We consider that the propagation in the 3D waveguide can be approximated by a 2D analogue where the monopoles are replaced by finite-size conducting scatterers, placed at the vertical positions of the antenna probes. Each scatterer radiates 2D cylindrical waves in the surrounding space and diffracts the energy coming from the environment. The field radiated from the emitter and the diffracted field from the scatterers interfere creating destructive or constructive patterns in the waveguide. By performing a full-wave numerical analysis via the commercial software  COMSOL Multiphysics \cite{comsol}, we calculated the total field in each position and frequency. The 2D approximation allows us to solve for large areas and frequency spans in a relatively short time and provides us with a qualitative evaluation of the propagation properties in a multiscattering environment. A priori, we assume that the antennas are impedance matched in all the spectrum of interest and that only the TEM mode is excited, both effectively controlled by the height of the structure. We investigate the system of 25$\times$25 nodes depicted in Fig. \ref{fig3}(a). Each antenna (scatterer) is a finite size copper cylinder of radius $R$ = 0.12 mm. In this approximation we do not take into account the impedance characteristics of the antennas. The emitter is simulated as a field source that radiates omnidirectional electromagnetic waves. All the surrounding scatterers reflect the incoming wave. In this way we estimate the energy profile of the propagating waves in the presence of the reflecting obstacles. Fig. \ref{fig3}(b,c) present  the profile of the total energy at frequency $f=$ 25 GHz, $f=$ 60 GHz and $f=$ 180 GHz when the emitter is no.1 and no.13, respectively.  Evidently, the electromagnetic waves interfere either destructively or constructively producing patterns of high or low energy corresponding to the dark and bright spots.  In the position of the receiver we also estimate the power captured by the multipath propagation coming from all  directions. The total power accumulated in the position of the receiver $M$ when $N$ emits, $P_{MN}$, is normalized by the total radiated power from the emitter $P_0$. The system is reciprocal, that is, $S_{MN} = S_{NM}$. Fig. \ref{fig3}(d) presents the power received in the position $M$ transmitted from emitter $N$ over the frequency range of $f$ =[25 GHz, 200 GHz] for node  pairs  schematically depicted in the insets. As observed in all cases, the received power remains on average the same for each pair in the entire frequency span. However, for nearly all cases, there are some frequency points where the received power drops. For example, for the case of the pair no7-no.17 (panel vi) there appear three dips in the received power at around $f=$ 45 GHz, $f=$ 80 GHz and $f=$ 115 GHz. These points correspond to destructive wave interference. Moreover we can observe the general tendency of the decreased received power with respect to the node-pair distance,i.e., for the pair no.1-no.21 (panel i) the average received power is -15 dB whereas for the pair no.1-no.6 (panel ii), the received power is on average -8 dB. Using this 2D qualitative analysis as a guideline, we can select the operation frequency for the actual 3D implementation of the wireless communication channel in the software-defined HSF.

\section{\label{sec:conc}Conclusion}
\vspace{-0.1cm}
In conclusion, we have addressed the issue of intercell wireless communication in the complex environment of a functional, software-defined metasurface. We have focused on two different scenarios with the communication taking place either in the metasurface plane or inside a dedicated channel. In both cases, we have assessed the performance by evaluating the electromagnetic field in the structure and calculating the scattering parameters between transmitting and receiving antennas. After careful design, we have obtained a transmission efficiency of -15dB and -8dB for scenarios A and B, respectively. We have thus demonstrated efficient wireless intercell connectivity without interfering with the metasurface operation taking an essential step towards realizing adaptive hypersurfaces with fully reconfigurable functionalities.


\section*{Acknowledgment}
\vspace{-0.1cm}
This work was supported by the European Union’s Horizon 2020 research and innovation programme-Future Emerging Topics (FETOPEN) under grant agreement No 736876.

\newpage

\end{document}